PAPER • OPEN ACCESS

# The effect of convection on the position of the free liquid surface under zero and terrestrial gravity

To cite this article: A I Fedyushkin 2020 *J. Phys.: Conf. Ser.* **1675** 012039

View the article online for updates and enhancements.

## You may also like

- Amount of Free Liquid Electrolyte in Commercial Large Format Prismatic Li-Ion Battery Cells
  Natalia P. Lebedeva, Franco Di Persio, Theodora Kosmidou et al.

- Breakup of free liquid jets influenced by external mechanical vibrations
  V N Lad and Z V P Murthy

- Ions at hydrophobic interfaces
  Yan Levin and Alexandre P dos Santos

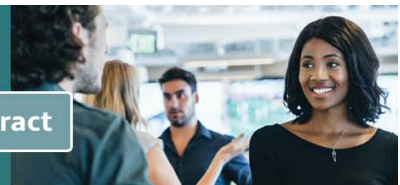
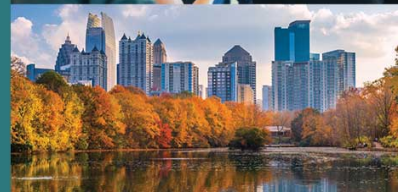

# The effect of convection on the position ofthe free liquid surface under zero and terrestrial gravity

**A I Fedyushkin**

Ishlinsky Institute for Problems in Mechanics of the Russian Academy of Sciences

E-mail: fai@ipmnet.ru

**Abstract.** Theresults of numerical modelinghas based on the solution of Navier-Stokes equations for convection oftwo-layer system air-water are presented.The influence of thermo-capillary and gravitational convection on the deformation and time averaged position of the interface of air-water system under sudden lateral heating of a square-section cavity are shown.

## 1. Introduction

A volume of liquid with a free surface can have different types and forms of interface: equilibrium, non-equilibrium, stable, unstable, stationary and non-stationary.The shape and type of free surface depend on the environmental conditions and properties of the liquid [1].The equilibrium position of the liquid volume corresponds to the minimum potential energyand free surface area reduces to possible minimum under the action of intermolecular interaction forces [1, 2].The surface tension forces of the liquid in interaction with the forces of pressure and viscosity create an energy-efficient form of the liquid volume.The shape of a liquid in a limited volume can be affected by the contact wetting angles,the ratio of liquid/gas properties and volumes [1, 3, 4]. The influence of surface forces on the shape of the interface is usually stronger in weightlessness than in the gravitational field of the Earth [1, 3-6].

The impermanence of the surface tension and (or) curvature of the free gas-liquid interface (or liquid-liquid) can cause the movement of liquids and gas along the interface [1-5].In this case, the liquid moves under the action of capillary forces, and the flow is directed from a low to a higher value of surface tension. This flow is called capillary convection or Marangoni convection. Surface tension impermanence can be caused by temperature and/or concentration gradients, in particular, by the presence of surfactants [3, 5, 7].The occurrence of Marangoni convection can be monotonous or threshold character.The appearance of a monotonous or threshold character of Marangoni convection can be caused by various reasons: the directions of heat and (or) impurity concentration gradients [4]; rheological properties of the liquid [7]; ambient conditions; the presence of wetting solids and their contact properties [1, 3, 4].The influences of thermal and concentration Marangoni convection on the deformation and stability of the free surface are described in numerous monographs and papers, for example, [1-3, 7, 8, 9].In paper [7] has been found that at sufficiently large Marangoni numbers (Ma ≥ 50 000) the diffusion process gives rise to instability in the system of immiscible liquids and a soluble surfactant, provided that their densities are equal, even in the absence of contraction.

The study of free surface behavior is important for many applications, for example, for aviation and cosmonautics (the behavior of fuel in the tanks of aircraft and missiles in zero gravity), the chemical industry, for firefighting technologies for obtaining materials and medical preparations, etc.[4, 6, 10, 11].

The results of this paper show that the position of the air-water interface (initially located parallel to the heat gradient) in square cavity half filled with water and lateral heated is not stable in zero gravity. The results of numerical simulation of the fact that in weightlessness the air-water interface can change its position by 90 degrees, aiming to take a stable average position (along isotherm) are demonstrated.

## 2. Problem statement and mathematical model

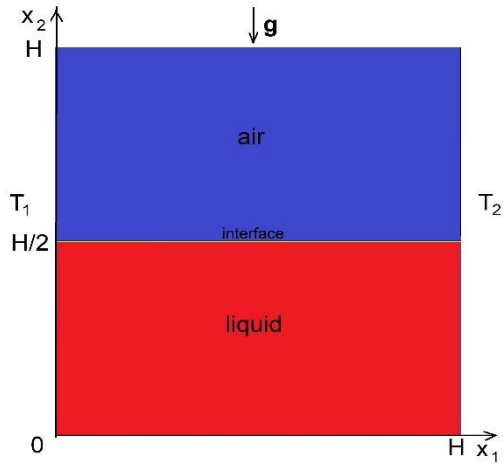

**Figure 1.** Scheme of computational domain and initial distribution of phases.

The problem of convection in a two-layer air-water system in a square cavity with slippage borders is considered (figure. 1).The influence of thermo-capillary (acceleration of gravity $\mathbf{g} = 0$) and gravitational convection ($\mathbf{g} \neq 0$) on the location of the air-water interface is analyzed.

At the initial moment, the interface is flat and horizontal. Water occupies the lower half of the calculated domain up to a height of H/2, as shown in figure 1.

The mathematical model is based on solving the system of two-dimensional Navier-Stokes equations for a two-phase gas-liquid system in the approximation of the "mixture" model [2] can be written in the form:

$$\frac{\partial u_1}{\partial x_1} + \frac{\partial u_2}{\partial x_2} = 0 \quad (1)$$

$$\frac{d(\rho u_1)}{dt} = -\frac{\partial p}{\partial x_1} + \frac{\partial}{\partial x_1}\left(\mu \frac{\partial u_1}{\partial x_1}\right) + \frac{\partial}{\partial x_2}\left(\mu \frac{\partial u_2}{\partial x_2}\right) + F_1 \quad (2)$$

$$\frac{d(\rho u_2)}{dt} = -\frac{\partial p}{\partial x_2} + \frac{\partial}{\partial x_1}\left(\mu \frac{\partial u_1}{\partial x_1}\right) + \frac{\partial}{\partial x_2}\left(\mu \frac{\partial u_2}{\partial x_2}\right) + F_2 - g_2 \beta_T (T - T_1) \quad (3)$$

$$\frac{\partial T}{\partial t} + u_1 \frac{\partial T}{\partial x_1} + u_2 \frac{\partial T}{\partial x_2} = a \left(\frac{\partial^2 T}{\partial x_1^2} + \frac{\partial^2 T}{\partial^2 x_2}\right) \quad (4)$$

where $df/dt = \partial f/\partial t + u_1 \partial f/\partial x_1 + u_2 \partial f/\partial x_2$ is the substantial derivative; $x_1$, $x_2$ are Cartesian coordinates; $u_1$, $u_2$ are velocity vector $\mathbf{u}(u_1, u_2)$ components, t is the time, p is the pressure; $\rho$ is the density; T is temperature; $\mathbf{g}(g_1, g_2)$ is gravitational acceleration of the Earth's free fall; $\beta_T$, a, $\mu$ are temperature expansion of the liquid, thermal diffusivity, the dynamic viscosity coefficients; $F_1, F_2$ are components of external force $\mathbf{F}(F_1, F_2)$, operating in a narrow zone along the air-liquid interface.

To describe the two-phase air-liquid system, we used the system of equations (1-3) with one equation for momentum transfer under the assumption of a "mixture" model [2] with averaged velocities $\mathbf{u} = \varepsilon \mathbf{u}_{air} + (1-\varepsilon)\mathbf{u}_{liquid}$, density $\rho = \varepsilon \rho_{air} + (1-\varepsilon)\rho_{liquid}$, and viscosity $\mu = \varepsilon \mu_{air} + (1-\varepsilon)\mu_{liquid}$, where the values with the index air refer to air, and with the index liquid refer to liquid. The volume fraction of liquid $\varepsilon$ ($0 < \varepsilon < 1$) was determined from the solution of the transfer equation: $\partial \varepsilon / \partial t + u_i \partial \varepsilon / \partial x_i = 0$.

The boundary conditions at the air–liquid interface were determined from the equilibrium condition of surface forces and pressure [2]:

$$(p_1 - p_2 + \sigma k)n_i = (\tau_{1ij} - \tau_{2ij})n_j + \partial \sigma / \partial x_i \qquad (5)$$

where $\tau_{\alpha ij} = \mu_\alpha \left( \dfrac{\partial u_i}{\partial x_j} + \dfrac{\partial u_j}{\partial x_i} \right)_\alpha$ is the viscous stress tensor $(i=1,2; j=1,2)$; index $\alpha$ denotes: α=1 – liquid, α=2 – air; $\sigma = \sigma(T)$ is the surface tension coefficient, which is a function of temperature and coordinates; $p_1$, $p_2$ are the fluid and air pressures; $\kappa = 1/R_1 + 1/R_2$ is the surface curvature where $R_1, R_2$ are the radii of curvature for liquid and air; **n** is the unit normal vector directed into the second fluid;

　　The conditions for the absence of friction are set at the boundaries of the computational domain and condition (5) at the interface of the two-phase liquid-air system. The modeling of the change in the shape of the air-liquid interface was performed using the model of liquid volumes (VOF - Volume Of Fluid method). The interface was determined by the VOF method with increased resolution and taking into account surface forces by the CSF method (Continuum Surface Force) [12]. The CSF method allows one to remove the singularity in the case of turning the radius of curvature of the interface surface to zero and increase the accuracy of the calculations [12]. When solving the problem, condition (5) at the interphase boundary was taken into account through an additional local bulk force F on the right side of the momentum transfer equation (2-3). The force F acts only in a very narrow area, enclosed along an interface curve $l$ of width $\Delta h$, where $l$ is the curve which shows the exact location of the air-water interface in the 2D calculation area. When $\Delta h$ tends to zero, **F** force can be written as $\mathbf{F}(l_s) = \sigma \kappa(l_s)\mathbf{n}(l_s)$ at each point $l_s$ of the interface curve $l$, where $\kappa(l_s)$ is curvature of curve $l$ at point $l_s$, $\mathbf{n}(l_s)$ is the normal at point $l_s$ of interface curve $l$ [12].

　　For numerical solution of the system of equations (1-4), the conservative method of control volumes with the approximation of spatial derivatives of the second order and first order in time was used [13]. The accuracy of defining the interface is limited by the size of the grid cells and the solution methods, so a detailed dynamic grid was used on both sides of the interface. The additional details, validation results of the mathematical model and the examples of its using are given in papers [14, 15].

　　The problem is characterized by geometric parameters, relative values of the properties of the two-layer system and the following dimensionless numbers: Marangoni $\mathrm{Ma} = -\dfrac{\partial \sigma}{\partial T} \dfrac{H \Delta T}{\nu \rho a}$, Rayleigh $\mathrm{Ra} = g\beta_T \Delta T H^3 / \nu a$, Prandtl $\mathrm{Pr} = \nu / a$, where, $\nu = \mu/\rho$ is the kinematic viscosity coefficient, H is the square size, $\Delta T = T_2 - T_1$ is the temperature scale ($T_2 > T_1$).

## 3. Results

At first, we will consider the case of thermo-capillary convection only for weightlessness (g=0) for a two-layer water-air system, as shown in figure 1. At the initial moment, the temperature was the same in whole the calculated domain. When the temperature difference between the vertical walls instantly will be change, thermo-capillary convection will arise at the interface. The liquid and air begin move in its parts of domain and the interface bends slightly and fluctuates over time, as a result of strong capillary convection with large Marangoni number ($\mathrm{Ma} = 10^6$). This oscillatory character of thermo-capillary convection can continue for a long time. Air and water are heated to a certain quasi-stationary temperature distribution, according to the specified boundary thermal conditions. The interface remains on average horizontal, as shown in figure 2, but due to changes in the temperature field throughout the volume, it becomes more unstable. The position of the air-water interface is shown in color in figure 2a, where water indicated by red and air indicated by blue. The temperature field is shown by colored contours in figure 2b (high temperature is shown by red color, low temperature is shown by blue) at time t=5 seconds ($\mathrm{Ma} = 10^6$, $\mathrm{Ra} = 0$). In figure 3 the time dependencies of the average ($h_{average}$), maximum ($h_{max}$) and minimum ($h_{min}$) deviations of the interface from the initial horizontal position ($Ma = 10^6$, $Ra = 0$) are presented.

It should be noted that this vibrational mode is unstable and since there is no gravity (and no stress on all the side walls of the region), the interface can change location at the slightest perturbations. The unstable quasi-equilibrium location of the air-water interface can be changed by thermo-capillary convection itself, but over a long period of time. For faster removal of the interface from an unstable quasi-equilibrium state, special short-term small perturbations of the flow of liquid and air can be created, for example, a single rotational oscillation. As a result, the location of water and air in the volume will be redistributed, and the temperature field in the area will also be redistributed. There will be damped oscillations of the air-water interface near the average vertical location. Thus, this will lead to an oscillatory process in the form of water splashes and to a new average location of the interface with predominantly conditionally vertical direction (along the $0x_2$ axis).

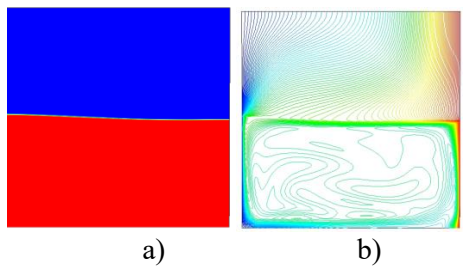
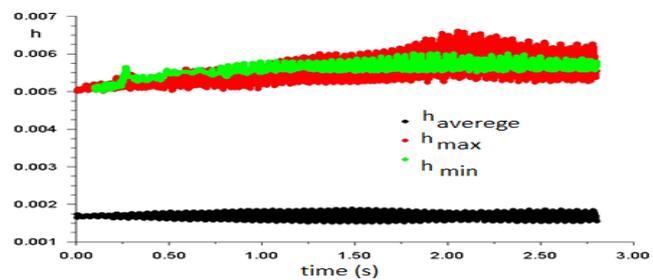

**Figure 2.** Position of the air-water interface (a) and the isotherm (b), for time t=5 sec, $Ma = 10^6$, $Ra = 0$.

**Figure 3.** Time dependencies of the average, maximum and minimum deviations of the interface from the initial horizontal position ($Ma = 10^6$, $Ra = 0$).

Instantaneous locations of air-water interface (on the left side of figures; water colored by red, air colored by blue) and the isotherm (on the right side; high temperature colored by red, low temperature colored by blue) after loss of initial horizontal interface location at four time moments: (a) - is t=11.89, (b) - is t=12.3, (c) - is t=15.86, (d) - is t=16.4 seconds are illustrated in figures 4.

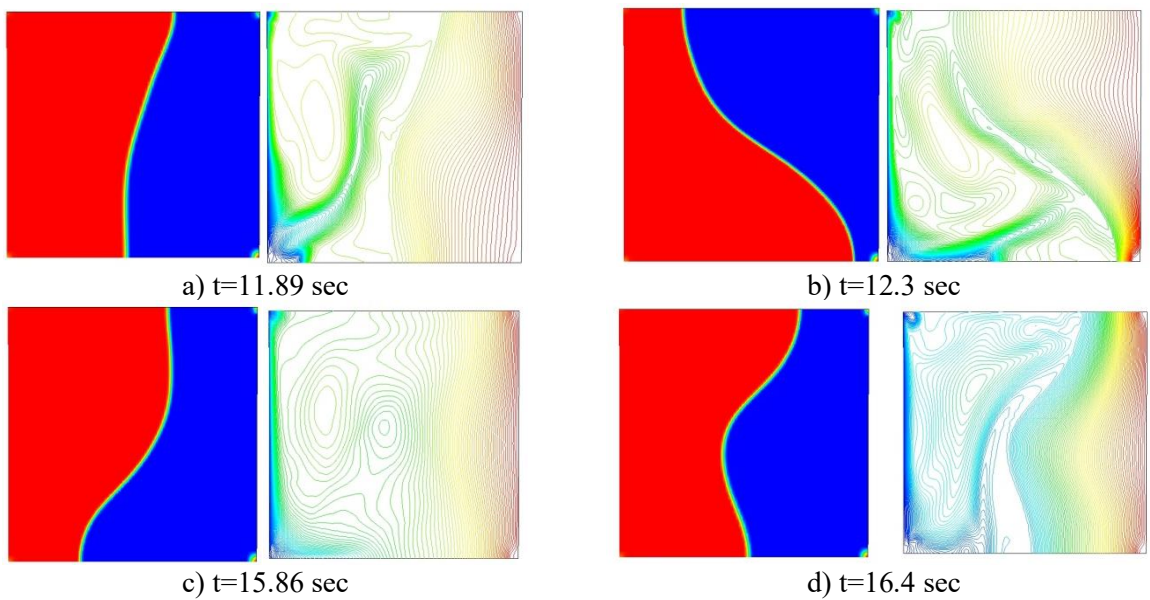

a) t=11.89 sec   b) t=12.3 sec
c) t=15.86 sec   d) t=16.4 sec

**Figure 4.** The instantaneous position of the interface (on the left) and isotherms (on the right) for different moments of time ($Ma = 10^6$, Ra=0).

It can be seen that the interface, having lost stability near the conditionally horizontal position, fluctuates near the conditionally vertical one. This is due to the fact that the location of the phase interface and the distributions of temperature in the phases are in nonlinear interaction. These interactions have different time scales of reaction on changes in temperature distribution and phase location. The Marangoni convective flow tends to stabilize fluctuations and increase the stability of the interface location.

After heating two-layer air-water system, temperature distribution in it becomes such that the isotherms acquire predominantly vertical direction (averaged over time), and the interface, turning at 90 degrees, mainly tends to occupy energetically minimum favorable position (conditionally vertical), adapting along the isotherms. This is characterized by the time-average fields of temperature and the positions of the interface and is confirmed by the simulation results. On the right in figures 5 color contours of temperature (red color is high temperature, blue is low) and on the left in figures 5 contours of the volume fraction (red color is water, blue is air) averaged over time in the range from t=0 to t=15.86 seconds for $Ma = 10^6$ and $Ra = 0$ are shown. The air-water interface change its location but performs high-amplitude oscillations near the conditionally vertical direction this can be seen in figure 6a. In figure 6 maximal range of deviations from the average location of the interface averaged over time in the range from t=0 to t=5 sec for $Ma = 10^6$ ((a) - is $Ra = 0$, (b) - is $Ra = 10^7$) are demonstrated by color (red is high, blue is low).

The gravitational convection (with an flow intensity corresponding to Rayleigh number of $Ra = 10^7$ or more) without thermo-capillary convection ($Ma = 0$) and in the presence of thermo-capillary convection ($Ma = 10^6$) is a stabilizing factor for the location of the free boundary (figure 6b). Therefore, in the presence of gravitational convection the interface remains mostly horizontal (heavy water there is at the bottom of domain below light air).

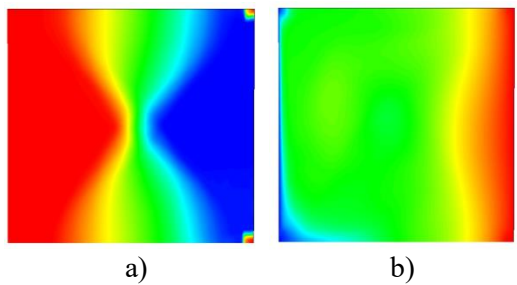
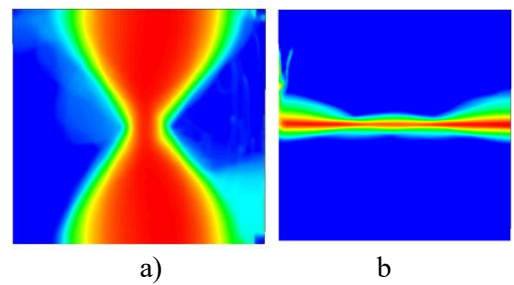

**Figure 5.** Contours of the volume fraction of water (a) and the isotherms (b) averaged over time across the range from t=0 to t=15.86 sec ($Ma = 10^6$, $Ra = 0$).

**Figure 6.** Contours of deviations from the average position of the interface averaged over time across the range from t=0 to t=5 sec ($Ma = 10^6$, (a) - $Ra = 0$, (b) - $Ra = 10^7$).

In weightlessness condition initial conditionally horizontal location of the phase interface is unstable at large Marangoni numbers. Thus, from the simulation results (figures 5, 6), it can be concluded that in the conditions of weightlessness after the loss of stability, the initial conditionally horizontal position of the air-water interface changes on time average to a quasi-stable one (the interface turns at 90 degrees and is mainly to locate along the isotherm conditionally in a vertical direction).

This effect is not present for terrestrial conditions because gravitational convection has a damping effect on the position of the free surface. The position of the interface under the Earth's conditions with the direction of the gravity acceleration vector parallel to the heated side walls remains on average over timehorizontal position.

**Conclusions**

The location of the phase interface (air-water) in weightlessness for large Marangoni numbers can be unstable when the interface is located parallel to the heat gradient. In weightlessness the interface of two-layer system laterally heated in a square cavity with walls without friction can change orientation its position by thermo-capillary convection and take a stable (energy-efficient) location(averaged over time) with parallel direction to the heated wall.After heating horizontal air-water system, temperature distribution in it becomes such that the isotherms acquire predominantly vertical direction (averaged over time), and the interface, turning at 90 degrees, mainly tends to occupy energetically minimum favorable direction of position (conditionally vertical), adapting along the isotherms.

Gravitational convection with and without thermo-capillary convection is a stabilizing factor for the horizontal position of the free boundary

**Acknowledgements**

The study was supported by the Government program (contract # AAAA-A20-120011690131-7) and was funded by RFBR, project number20-04-60128.